\documentclass[fleqn,twoside,twocolumn,nofootinbib,showkeys]{revtex4} 
\usepackage[nocpr]{ujp} 
\usepackage[utf8]{inputenc}
\usepackage[english]{babel}

\usepackage[usenames]{color}

\newcommand{\Hc}{\mathcal{H}}
\newcommand{\Fc}{\mathcal{F}}

\begin{document}

\title[Vacuum birefringence in the field of a current loop and a guided electromagnetic wave]%
{Vacuum birefringence in the field of a current loop and a guided electromagnetic wave}

\author{O. Novak}
\affiliation{The Institute of Applied Physics of National Academy of Sciences of Ukraine}
\address{Petropavlivska str. 58, 40000 Sumy, Ukraine}
\email{novak-o-p@ukr.net}

\author{M. Diachenko}
\affiliation{The Institute of Applied Physics of National Academy of Sciences of Ukraine}
\address{Petropavlivska str. 58, 40000 Sumy, Ukraine}
\email{mykhailo.m.diachenko@gmail.com}

\author{E. Padusenko}
\affiliation{The Institute of Applied Physics of National Academy of Sciences of Ukraine}
\address{Petropavlivska str. 58, 40000 Sumy, Ukraine}
\email{elenlevitskaya@gmail.com}

\author{R. Kholodov}
\affiliation{The Institute of Applied Physics of National Academy of Sciences of Ukraine}
\address{Petropavlivska str. 58, 40000 Sumy, Ukraine}
\email{kholodovroman@yahoo.com}

\udk{530.145} 
\razd{\secix}

\autorcol{O.\hspace*{0.7mm}Novak, M.\hspace*{0.7mm}Diachenko, E.\hspace*{0.7mm}Padusenko, R.\hspace*{0.7mm}Kholodov}%

\begin{abstract}
The effect of vacuum birefringence in a magnetic field generated by a laser-driven capacitor-coil generator as well as in field of an electromagnetic wave in a radioguide has been theoretically considered. 
The resulting ellipticity of a linearly polarized laser beam propagating in the considered field configuration is calculated.
The obtained results are compared to the parameters of the PVLAS experiment which aims to observe the effect of magnetic vacuum birefringence experimentally.\end{abstract}


\setcounter{page}{1}%

\maketitle

\section{Introduction}
\label{sec:intro}
\begin{figure}
  \vskip1mm
  \includegraphics[width=0.7\columnwidth]{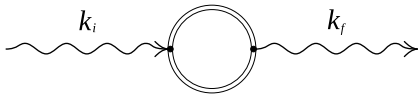}
  \vskip-3mm
  \caption{%
  Feynman diagram of photon polarization flip in a background magnetic field. 
  The double lines represent the electron propagator in external field.
  }
  \label{fig:loop}
\end{figure}

The equations of classical electrodynamics are linear, consequently the classical theory does not consider the photon-photon interaction as well as photon-external-field interaction. 
The uncertainty relation, however, allows production of virtual electron-positron pairs by photons. 
Such pairs may interact with fields and other photons. 
Thus, the quantum theory involves a number of nonlinear vacuum effects; variation of the polarization of photons propagating in an external magnetic field is among them. 
This effect, known as the magnetic vacuum birefringence, was first predicted in \cite{Heisenberg36}.
Within the framework of the scattering theory, this process is described by the Feynman diagram shown in Fig.~\ref{fig:loop}.
Studies of this process  requires the computation of the polarization operator generally~\cite{Shabad75, Shabad88, Shabad04, Shabad10}.
The Euler-Heisenberg effective Lagrangian obtained in~\cite{Heisenberg36} is widely used in low-energy approximation, $\hbar \omega \ll mc^2$.

The first direct evidence of vacuum birefringence is the high polarization degree of the optical radiation of an isolated neutron star 
RX J1856.5--3754~\cite{Mignani17}. 
According to the obtained models of the pulsar atmospheres, the considerable polarization of their radiation can be explained only by vacuum effects in superstrong fields of neutron stars~\cite{Heyl00, Heyl02, Adelsberg06, Fernandez11, Taverna15, Gonzalez16}.

Under laboratory conditions, the effect has not yet been observed due to its weakness in the fields available for the experiment. 
Direct observation of magnetic vacuum birefringence is the aim of the PVLAS (Polarizzazione del Vuoto con LASer) experiment, which is held at the National Institute of Nuclear Physics in Ferrara, Italy~\cite{Zavattini08, DellaValle14, DellaValle16}.
The technique is based on measuring of the magnitude of the ellipticity acquired by a linearly polarized laser beam with a wavelength  $\lambda = 1064$~nm when passing a distance $L = 1,64$~m in a magnetic field with an induction $B \approx 2.5$ in vacuum. 
To enhance the effect, a Fabry-Pierrot resonator is used that increases the path of the beam in the field by a factor of $\sim 10^5$.
For the given parameters, the theoretical value of the ellipticity is $\psi \approx 1.6\cdot 10^{-16}$ after a single pass.
Despite significant progress in experimental techniques, the accuracy of measuring the ellipticity is still an order of magnitude lower than that required to observe the effect~\cite{DellaValle16}.

In this paper, vacuum birefringence in alternative field configurations is theoretically studied. 
Section~\ref{sec:theory} contains a brief overview of the theory of the vacuum birefringence effect. 
In Section~\ref{sec:coil}, we calculate the ellipticity asquired by a linearly polarized beam in the field generated with laser-driven capacitor-coil target.
In Section~\ref{sec:rg}, the possibility of observing the effect using radio frequency waveguides is discussed.

\section{Theoretical background}
\label{sec:theory}

To study the evolution of the polarization of a beam in an external field, it is convenient to use the technique developed in~\cite{Shakeri17}.
The relativistic system of units  is used through this section, $\hbar = c = 1$.

Photon polarization is normally described in terms of the Stokes parameters $Q,U,V$. 
These parameters are closely related to the density matrix of probability of photons.
In the basis of linear polarizations the given matrix has the following form,
\begin{equation}
\label{rho}
    \rho = \frac12
    \begin{pmatrix}
        I+Q & U-iV\\
        U+iV & I-Q
    \end{pmatrix}.
\end{equation}
Thus, the evolution of the Stokes parameters can be found if the equation for $\rho$ is known.
Note that $\rho$ can be expressed via the average of the photon number operator $\hat D_{ij} = \hat a_i^+ \hat a_j$,
\begin{equation}
    \langle D_{ij}(\vec k)\rangle = (2\pi)^{3} 2\omega \delta^3(0) \rho_{ij}(\vec k).
\end{equation}
where $\omega$ and $\vec k$ are the photon frequency and the wave vector. 
In turn, the evolution of the photon number operator is determined by the interaction Hamiltonian~\cite{Kosowsky96},
\begin{equation}
\label{dDdt}
    \left< \frac{d\hat{D}_{ij}}{dt} \right>    \approx 
    i \left<\left[ \hat{\Hc}_{int}, \hat{D}_{ij} \right]\right>.
\end{equation}
This expression was obtained in the lowest approximation of the interaction of photons with the field $\hat{\Hc}_{int}$. 
The right-hand side of (\ref{dDdt}) describes only the variation of the photon polarization, whereas the terms corresponding to the scattering processes were neglected.

In the low energy approximation, it is advisable to use the effective Euler-Heisenberg Lagrangian~\cite{Heisenberg36},
\begin{equation}
\label{Lint}
    \mathcal{L}_{int} = - \frac{\alpha^2/(4\pi)^2}{180m^4} \left[ 
        5(\Fc_{\mu\nu}\Fc^{\mu\nu})^2  - 14\Fc_{\mu\nu}\Fc^{\nu\lambda}\Fc_{\lambda\sigma}\Fc^{\sigma\mu}
    \right],
\end{equation}
where $\alpha$ is the fine structure constant and $m$ is the electron mass.

In order to consider the interaction of photons with an external field, we replace the electromagnetic field tensor in (\ref{Lint}) in accordance with  in  $\Fc^{\mu\nu} = F^{\mu\nu} + f^{\mu\nu}$, where  $F^{\mu\nu}$ is a slowly varying external field, and the quantum field $f^{\mu\nu}$ plays the role of a small perturbation to the Hamiltonian.

To use the equation (\ref{dDdt}), the tensor $f^{\mu\nu}$ has to be expressed in terms of operators of the photon creation and annihilation $\hat a$ and $\hat a^+$, entering the 4-potential $A_\nu$:
\begin{equation}
    f^{\mu\nu} = \partial_\mu \hat{A}_\nu - \partial_\nu \hat{A}_\mu,
\end{equation}
\begin{equation}
\label{Anu}
    \hat{A}^\nu = \sqrt{4\pi} \int \frac{d^3k/(2\pi)^3}{\sqrt{2\omega}} 
    \sum\limits_{s = 1,2}
    \left[
        \hat a_{\vec k s}e^\nu_s \mbox{e}^{-ikr} + \hat a^+_{\vec k s}e^{\nu *}_s \mbox{e}^{ikr}
    \right],
\end{equation}
where $e_{1,2} = (0, \vec e_{1,2})$ are the 4-vectors of the photon polarization.

Using expressions (\ref{rho})--(\ref{Anu}) and performing simple transformations we obtain a system of differential equations for the Stokes parameters. 
This system can be conveniently written in the vector form by introducing the Stokes vector $\vec{S} = (Q,U,V)$:
\begin{equation}
\label{SOmega}
    \dot{\vec{S}} = \left[\vec{\Omega} \times \vec{S} \right],
\end{equation}
\begin{equation}
\label{Omega}
    \begin{array}{l}
        \Omega_x = -G \left(\varepsilon_2^2 - \varepsilon_1^2\right), \\
        \Omega_y = 2G\varepsilon_1\varepsilon_2, \\
        \Omega_z = 0.
    \end{array}
\end{equation}
Here, the following notation is used: 
%
\begin{equation}
\label{Gconst} 
    G = \frac{\alpha^2}{4\pi}\frac{2\omega}{15m^4}
\end{equation}
\begin{equation}
\label{eps}
    \varepsilon_i = \vec{E}\vec{e}_i - \vec{B} \left[\vec{n}\times \vec{e}_i \right], \qquad i = x,y,
\end{equation}
where $\vec E$ and $\vec B$ are the strength and induction of the external field,
$\vec{n} = \vec{k}/k$ is the unit vector defining the direction of the wave. 
The equations (\ref{SOmega}) was also obtained in \cite{Kubo80} and analyzed in detail for a constant uniform external field~\cite{Kubo83, Kubo85}. 
The numerical solution of Eq.~(\ref{SOmega}) allows one to find the polarization of photons in a slowly varying electromagnetic field of arbitrary configuration.

\section{Laser-driven capacitor-coil generator}
\label{sec:coil}
As it follows from equations (\ref{SOmega})--(\ref{eps}), the magnitude of the vacuum birefringence is proportional to the squared induction of a background magnetic field. 
Thus, a natural approach to experimental observation of the effect is increasing the field strength. 
Under laboratory conditions, the strongest stationary field with the induction of about 45~T was obtained using hybrid superconducting magnets~\cite{Schneider04}. 
Non-destructive methods of generation of a pulsed field make it possible to produce fields of about 100~T and duration of several microseconds. 
Recently, considerable interest has arised in laser-driven capacitor-coil targets able to produce field pulses of about several hundred T and pulse duration of the order of nanoseconds~\cite{Daido86, Daido87, Zhivopistsev91, Courtois05, Liao16, Law16}.

\begin{figure}
  \vskip1mm
  \includegraphics[width=\column]{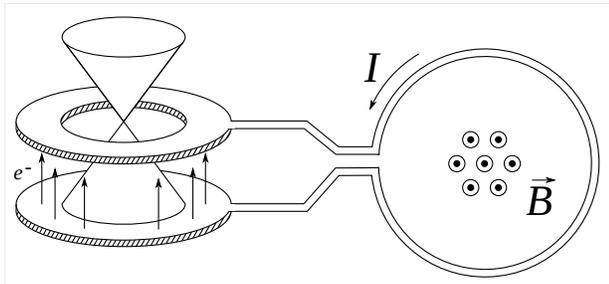}
  \vskip-3mm
  \caption{%
  Schemati diagram of a laser-driven capacitor-coil target.
  }
  \label{fig:coil-setup}
\end{figure}

The generator is composed of a single wire loop connected to the capacitor-target (Fig.\ref{fig:coil-setup}). 
One of the plates of the target is irradiated with a powerful laser pulse. 
Resonant absorption of radiation leads to a strong heating of electrons, which hit the other plate of the capacitor and produce a significant voltage between the plates. 
The resulting electric current in the coil produces a strong magnetic field. 
The observation of the peak magnetic induction of $610 \pm 30$~T in experiments with laser-driven coil with the coil diameter of 0.5 mm was reported in~\cite{Law16}.

Let us estimate the magnitude of the ellipticity caused by the vacuum birefringence  in a possible experiment with such field generator. 
For simplicity, we consider the quasistationary field. 
Such assumption is based on the fact that the beam passing time in this case is picoseconds, which is 3 orders of magnitude less than the field lifetime. 
We assume that the generator is a circular current loop and use the cylindrical coordinate system with the origin located at the center of the loop and the $z$-axis is directed perpendicular to the plane of the ring.
The radius of the loop is chosen to be $R = 0.5$~mm, and the magnetic induction at its center is $B_0 = 500$~T. 
The field components of the current loop in a point with coordinates $(\rho, \phi, z)$ is determined by the radius and induction at the center as
\begin{equation}
\label{coilH}
    \begin{array}{l}
        \displaystyle
        B_{\rho} =   \frac{B_0 R}{\pi} \frac{z}{u^2w\rho}
        \left[ (z^2+\rho^2+R^2)E(q^2) - u^2 K(q^2) \right], \\
        \displaystyle
        B_{\varphi} =  0, \\
        \displaystyle
        B_{z} = \frac{B_0R}{\pi}\frac{1}{u^2w} \left[
            (R^2-z^2-\rho^2)E(q^2) + u^2K(q^2)
        \right],
    \end{array}
\end{equation}
where
\begin{equation}
    \begin{array}{l}
        q^2 = 1 - u^2/w^2,\\
        u = z^2 + (R-\rho)^2, \\
        w = z^2 + (R+\rho)^2.
    \end{array}
\end{equation}

Equations (\ref{SOmega}) were numerically integrated in the external field (\ref{coilH}). 
The initial polarization of the laser beam was chosen to be linear with the Stokes parameters
\begin{equation}
\label{S0}
    \vec S_0 = (0, 1, 0),
\end{equation}
polarization plane forms the angle of 45$^\circ$ with the axes of the coordinate system where the polarization is determined.

As a result of traversing an external field, the linearly polarized beam acquires non-zero circular polarization, $V \neq 0$. 
In practice, it is convenient to characterize this polarization with the ellipticity $K$.
In the case when $U=1$ and $V \ll 1$, the ellipticity $K$ reads
\begin{equation}
\label{V2K}
    K \approx 2V.
\end{equation}

\begin{figure}
  \vskip1mm
  \includegraphics[width=\column]{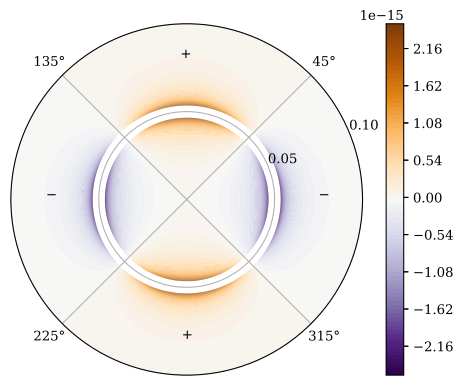}
  \vskip-3mm
  \caption{%
  The final ellipticity of beams propagating perpendicular to the coil plane.
  }
  \label{fig:coil-axial}
\end{figure}

Fig.~\ref{fig:coil-axial} depicts the resulting ellipticity of the beam propagating perpendicular to the ring plane.
The picture is symmetric with respect to rotation by 180$^\circ$. 
Indeed, equations (\ref{SOmega})--(\ref{eps}) are quadratic in magnetic field, therefore, they do not change when replacing $\vec B \rightarrow -\vec B$. 
It also follows from equations (\ref{eps}) that only the field components  perpendicular to the wave vector of the beam make a contribution. 
Indeed, as can be seen from fig.~\ref{fig:coil-axial} the effect is absent for when the beam propagates through the center of the loop. 
The greatest effect is observed near the loop itself, but this situation is unfavorable for experimental observation, since the loop evaporates during the current pulse, which will inevitably introduce uncertainty to the result.

\begin{figure}
  \vskip1mm
  \includegraphics[width=\column]{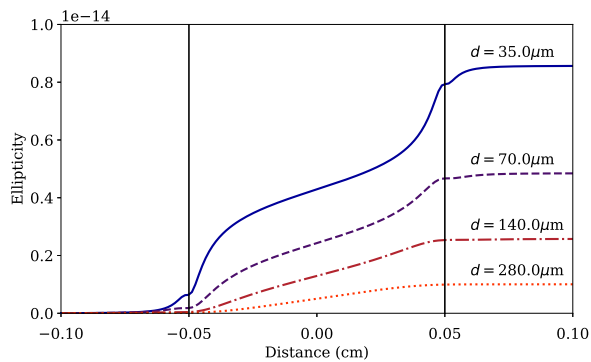}
  \vskip-3mm
  \caption{%
  The dependence of ellipticity on the distance passed in the field of a current loop. 
  The beam propagates parallel to the coil at a distance $d$ in the same plane with the loop center. 
  The vertical lines mark the position of the loop wires.
  }
  \label{fig:coil-evo}
\end{figure}

\begin{figure}
  \vskip1mm
  \includegraphics[width=\column]{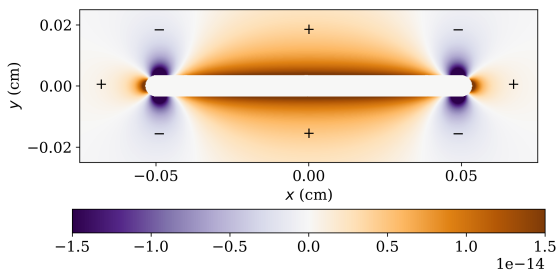}
  \vskip-3mm
  \caption{%
  The final ellipticity of beams propagating parallel to the coil plane.
  }
  \label{fig:coil-side}
\end{figure}

A more appropriate is a configuration with  the beam propagating parallel to the plane of the wire loop and at some distance $d$ from it. 
Fig.~\ref{fig:coil-side} depicts the final ellipticity of the rays in this case. 
Note that the effect can be enhanced by using two generators in the Helmholtz configuration, as has been done in~\cite{Law16}.
Fig.~\ref{fig:coil-evo} shows the evolution of ellipticity in this case for different values of $d$. 
As can be seen, the magnitude of the effect can reach $10^{-15}$, which is an order of magnitude greater than in the PVLAS experiment~\cite{DellaValle14, DellaValle16}. 

Note that in PVLAS a Fabry-Pierrot resonator with a multiplication factor of $N \sim 10^5$  is used to increase the beam path in the field \cite{DellaValle14, DellaValle16}. 
The pulsed character of the field generated by a laser-driven coil can limit the possible values of the multiplication factor. 
Indeed, suppose that the size of the interferometer is comparable to the size of the coil itself. Then the time of flight of the beam through the resonator is $\sim 10^{-12}$s. 
Taking into account that the field lifetime is several nanoseconds, we find that the maximum multiplication factor also reaches $10^5$. 

Thus, it can be concluded that usage of a laser-driven coil in experiments on the observation of vacuum birefringence can increase the effect by an order of magnitude. 
On the other hand, the destructive nature of the generator can complicate such experiments and reduce their accuracy.

Let us estimate the residual atmosphere pressure $p$, when the vacuum birefringence has the same magnitude as the similar effect in gases (the Cotton-Mouton effect). 
The ellipticity associated with the Cotton-Mouton effect can be found from the expression~\cite{Rizzo97, DellaValle16}:
\begin{equation}
\label{CM}
    K = \frac{\pi L}{\lambda} \left(\frac{B}{1 \mbox{Tesla}}\right)^2 
    \left(\frac{p}{1 \mbox{atm}}\right) \Delta n_u,
\end{equation}
where $\Delta n_u$ is the value at pressure of 1~atm and magnetic induction of 1~Tesla. 
Values of $\Delta n_u$ were measured experimentally~\cite{Rizzo97, Mei10}. 
For example, for argon we have $\Delta n_u \approx 4.3 \cdot 10^{-15}$. 
Then for the typical parameter values one can estimate the corresponding pressure as $p \sim 10^{-9}$~atm. 
Note that a similar estimation for the conditions of the PVLAS experiment gives the same pressure value.

\section{RF waveguide}
\label{sec:rg}

\begin{figure}
  \vskip1mm
  \includegraphics[width=\column]{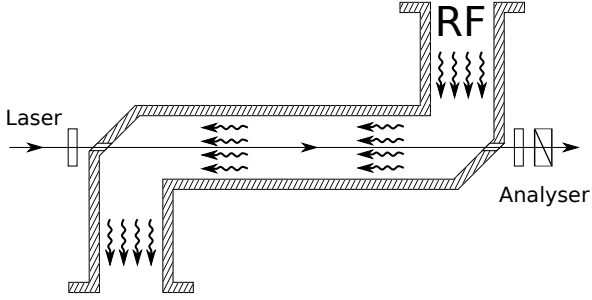}
  \vskip-3mm
  \caption{%
  Possible experimental setup with an RF waveguide as a source of a background field.
  }
  \label{fig:rg}
\end{figure}

Recently, considerable interest couses the possibility of observing the vacuum birefringence in the field of a laser wave. 
Modern laser systems can poduce field strengths that greatly exceed the values of the magnetic field available in laboratory conditions. 
As part of the Extreme Light Infrastructure (ELI) project, it is planned to produce even more powerful lasers, whose fields will reach the critical quantum electrodynamic value of $\sim 10^{16}$~V/cm.

In this paper, we consider an alternative experimental setup with an electromagnetic wave propagating in a rectangular hollow waveguide  as a background field~(Fig.~\ref{fig:rg}).
Currently, the most powerful sources of microwave radiation are klystrons. 
In particular, klystrons with an operating frequency of 11.424 GHz and peak power of 75~MW have been built at the SLAC National Accelerator Laboratory (SLAC National Accelerator Laboratory, USA)~\cite{Sprehn06}.

It is known~\cite{NikolskijEDRW}, that electromagnetic field in a waveguide has the form of standing waves in transverse directions and a traveling wave in the longitudinal direction,
\begin{equation}
\label{fexp}
    \vec F = \mbox{Re}\vec{\Fc} \mbox{e}^{i(\omega t - \Gamma z)}.
\end{equation}

Transverse wave numbers are determined by the waveguide geometry,
\begin{equation}
\label{chi}
    \begin{array}{l}
        \chi^2 = \chi_x^2  +  \chi_y^2, \\
        \chi_x = \frac{m\pi}{a}, \qquad 
        \chi_y = \frac{n\pi}{b},
    \end{array}
\end{equation}
where $a$ and $b$ are the waveguide sides, $m$ and $n$ are integer numbers. 
The longitudinal wave number $\Gamma$ is defined as
\begin{equation}
\label{gamma} 
    \Gamma^2 = \left(\frac{\omega}{c}\right)^2 - \chi^2.
\end{equation}
The frequency $\omega_c$ which satisfies the condition $\Gamma = 0$ is called the cut-off frequency. 
For frequencies below $\omega_c$, the longitudinal wave number becomes imaginary, and the wave propagation is impossible.

Waves propagating in hollow waveguides have non-zero longitudinal component of electric or magnetic field. 
Accordingly, possible waves belongs to H- or E-type. 
The numbers $m$ and $n$ defining the transverse wave number are commonly written as subscripts near the mode type designation. 
The waves H$_{01}$ and H$_{01}$ have the lowest frequency and are called the main modes.

The transverse components of the H-wave read
\begin{equation}
\label{Hfield}
    \left\{
    \begin{array}{l}
        E_x = -F_0 \frac{\chi_y}{\chi}     \cos\chi_xx\sin\chi_yy,\\
        E_y = -F_0 \frac{\chi_x}{\chi}     \sin\chi_xx\cos\chi_yy, \\
        B_x = -F_0 \frac{\chi_x}{\chi} W_E \sin\chi_xx\cos\chi_yy,\\
        B_y = -F_0 \frac{\chi_y}{\chi} W_E \cos\chi_xx\sin\chi_yy
    \end{array}
    \right.    
\end{equation}
where $W_E$ is wave impedance.

The amplitude of the field $F_0$ can be determined by the energy flux $\bar P$ through the waveguide crosssection,
\begin{equation}
\label{F0}
    F_0^2 = \bar P \frac{4\pi}{c} \frac{W_H}{ab}\delta_0,
\end{equation}
where $\delta_0 = 4$ for the fundamental wave and $\delta_0 = 8$ otherwise.
     
If a waveguide is not filled with media, the wave impedances $W_H$ and $W_E$ satisfy the relations
\begin{equation}
\label{WEH}
    W_E = \frac{\Gamma}{\omega/c},  \qquad W_EW_H = 1.
\end{equation}

The amplitude of the E-mode is given by similar expressions~\cite{NikolskijEDRW}.

Let us calculate the ellipticity asquired by a linearly polarized probe beam ($U=1$) when propagating in a waveguide parallel to its axis. 
For convenience of comparison, we choose the wavelength of the probe beam  $\lambda = 1064$~nm and the path length $L = 164$~cm, which corresponds to conditions of the PVLAS experiment. 
Parameters of the microwave source that are based on the desing of high-power klystrons developed at SLAC, namely,  power equals to 75~MW and frequency is $\nu = 11.424$~GHz. 
In order to fully define the background field (\ref{Hfield}), it is necessary to set geometrical dimensions of the waveguide. 
Note that the properties of the fundamental mode $H_{10}$ depend only on length of one side of waveguide, let it be $a$. 
As an illustration, we set the aspect ratio as $a:b = 3:1$, and the actual size $a$ is determined from the condition that the detuning of the operating frequency of the microwave from the cut-off frequency is
$(\omega-\omega_c)/\omega = 0.1$ .

Fig.~\ref{fig:rg} shows the asquired ellipticity of the probe beam at different points of the cross section of the waveguide. 
Apparently, the ellipticity magnitude is of order $10^{-18}$, which is significantly worse than in the PVLAS experiment with a constant magnetic field. 
The reason for this difference is the low value of the field strength in the waveguide, which in this case is only 0.2~T. 
In order to reach the value of 2.5~T, the power $\bar P$ should be of order of $\sim 20$~GW, which presently is not feasible.


\begin{figure}
  \vskip1mm
  \includegraphics[width=\column]{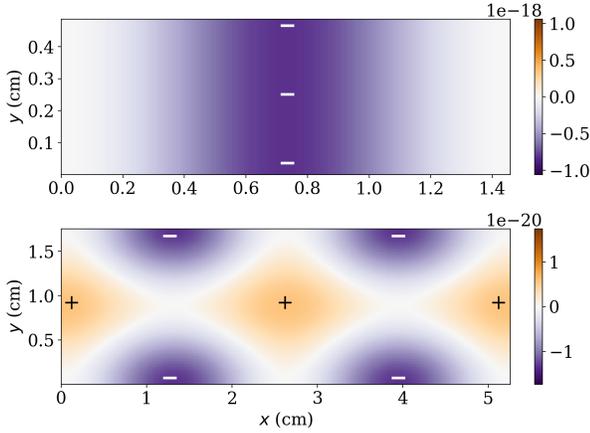}
  \vskip-3mm
  \caption{%
  The final ellipticity of beam propagating along the waveguide at different points of its cross section. 
  Upper panel: the case of the fundamental (H$_{10}$) mode; bottom panel: E$_{21}$ mode.
  }
  \label{fig:rg}
\end{figure}


\section{Conclusions}
\label{sec:conclusions}
The vacuum birefringence in an external field was theoretically predicted at the beginning of the twentieth century but has not yet been observed experimentally. 
To date, the most accurate experiment for measuring vacuum  birefringence is PVLAS~\cite{DellaValle14, DellaValle16}, which uses a magnetic field of 2.5~T as a background field. 
Much is expected of the future laser facilities that allow to achieve critical value of field strength.

In this paper, we consider the possibility of observing of the effect in alternative configurations of an external field, namely, in magnetic field of a laser driven coil and in the field of an RF waveguide.

In the case of a laser driven coil, the observed effect of vacuum birefringence may be an order of magnitude greater the corresponding value in the conditions of the PVLAS experiment. 
On the other hand, the destructive nature of laser-driven coils presents technical difficulties and reduce the accuracy of the experiment.

In the case of an RF-waveguide, the effect is much lower for existing microwave sources. 
The power required to increase the effect to the observable values is at least 15~GW, which greatly exceeds the current capabilities.

The research is supported by the grat of National Academy of Sciences of Ukraine for young scientists No~0117U001760.

\end{document}